\documentclass[]{tufte-handout}   	
\usepackage{geometry}                		
\usepackage{graphicx}				

\newcommand{\ci}[1]{$^{\mathrm #1}$}

\definecolor{gray}{rgb}{.5,0.5,0.5}

\title{A rocky planet transiting a nearby low-mass star}
\author{{\small Zachory K. Berta-Thompson$^{1,2}$, 
	Jonathan Irwin$^2$, 
	David Charbonneau$^2$, 
	Elisabeth R. Newton$^2$, 
	Jason A. Dittmann$^2$, 
	Nicola Astudillo-Defru$^3$, 
	Xavier Bonfils$^{4,5}$, 
	Michaël Gillon$^6$, 
	Emmanuël Jehin$^6$, 
	Antony A. Stark$^2$, 
	Brian Stalder$^7$, 
	Francois Bouchy$^{3,8}$, 
	Xavier Delfosse$^{4,5}$, 
	Thierry Forveille$^{4,5}$, 
	Christophe Lovis$^3$, 
	Michel Mayor$^3$, 
	Vasco Neves$^9$, 
	Francesco Pepe$^3$, 
	Nuno C. Santos$^{10,11}$, 
	Stéphane Udry$^3$ \& 
	Anaël Wünsche$^{4,5}$}}
\date{}							
\begin{document}

\renewcommand{\figurename}{{\bf Figure}}
\renewcommand{\thefigure}{{\bf \arabic{figure}}}
\setcounter{figure}{0}

\maketitle

	{\scriptsize \it
	\begin{itemize}
\setlength\itemsep{.2em}

\item[]		$^1$ Kavli Institute for Astrophysics and Space Research, Massachusetts Institute of Technology, \\77 Massachusetts Avenue, Cambridge, Massachusetts 02139, USA.
\item[]		$^2$ Harvard-Smithsonian Center for Astrophysics, \\60 Garden Street, Cambridge, Massachusetts 02138, USA.
\item[]		$^3$ Observatoire de Gen\`eve, Universit\'e de Gen\'eve, \\51 chemin des Maillettes, 1290 Sauverny, Switzerland.
\item[]		$^4$ Universit\'e Grenoble Alpes, IPAG, F-38000 Grenoble, France.
\item[]		$^5$ CNRS, IPAG, F-38000 Grenoble, France.
\item[]		$^6$ Institut d'Astrophysique et de G\'eophysique, Universit\'e de Li\`ege, \\All\'ee du 6 Ao\^ut 17, B\^atiment B5C, 4000 Li\`ege, Belgium.
\item[]		$^7$ Institute for Astronomy, University of Hawaii at Manoa, \\Honolulu, Hawaii 96822, USA.
\item[]		$^8$ Laboratoire d'Astrophysique de Marseille, UMR 6110 CNRS, Universit\'e de Provence, \\38 rue Fr\'ed\'eric Joliot-Curie, 13388, Marseille Cedex 13, France.
\item[]		$^9$ Departamento de F\'isica, Universidade Federal do Rio Grande do Norte, \\59072-970 Natal, Rio Grande do Norte, Brazil.
\item[]		$^{10}$ Instituto de Astrof\'isica e Ci\^encias do Espa\c{c}o, Universidade do Porto, \\CAUP, Rua das Estrelas, 4150-762 Porto, Portugal.
\item[]		$^{11}$ Departamento de F\'isica e Astronomia, Faculdade de Ci\^encias, Universidade do Porto, \\Rua Campo Alegre, 4169-007 Porto, Portugal. 
 
\end{itemize}
	}

{\bf 
M-dwarf stars -- hydrogen-burning stars that are smaller than 60 per cent of the size of the Sun -- are the most common class of star in our Galaxy and outnumber Sun-like stars by a ratio of 12:1. Recent results have shown that M dwarfs host Earth-sized planets in great numbers\ci{1,2}: the average number of M-dwarf planets that are between 0.5 to 1.5 times the size of Earth is at least 1.4 per star\ci{3}. The nearest such planets known to transit their star are 39 parsecs away\ci{4}, too distant for detailed follow-up observations to measure the planetary masses or to study their atmospheres. Here we report observations of GJ 1132b, a planet with a size of 1.2 Earth radii that is transiting a small star 12 parsecs away. Our Doppler mass measurement of GJ 1132b yields a density consistent with an Earth-like bulk composition, similar to the compositions of the six known exoplanets with masses less than six times that of the Earth and precisely measured densities\ci{5-11}. Receiving 19 times more stellar radiation than the Earth, the planet is too hot to be habitable but is cool enough to support a substantial atmosphere, one that has probably been considerably depleted of hydrogen. Because the host star is nearby and only 21 per cent the radius of the Sun, existing and upcoming telescopes will be able to observe the composition and dynamics of the planetary atmosphere.
}

We used the MEarth-South telescope array\ci{12} to monitor the brightness of the star GJ 1132, starting on 28 January 2014. The array consists of eight 40-cm robotic telescopes located at the Cerro Tololo Inter-American Observatory (CTIO) in Chile, and observes a sample of M-dwarf stars that are within 33 parsecs of Earth and smaller than 0.35 Solar radii. Since early 2014, the telescopes have gathered data almost every night that weather has permitted, following a strategy similar to that of the MEarth-North survey\ci{13}. On 10 May 2015, GJ 1132 was observed at 25-minute cadence until a real-time analysis system identified a slight dimming of the star indicative of a possible ongoing transit, and commanded the telescope to observe the star continuously at 0.75-minute cadence. These triggered observations confirmed the presence of a transit with a sharp egress (Fig. 1). 
\vfill
\begin{figure}
   \centering
   \includegraphics[width=4.125in]{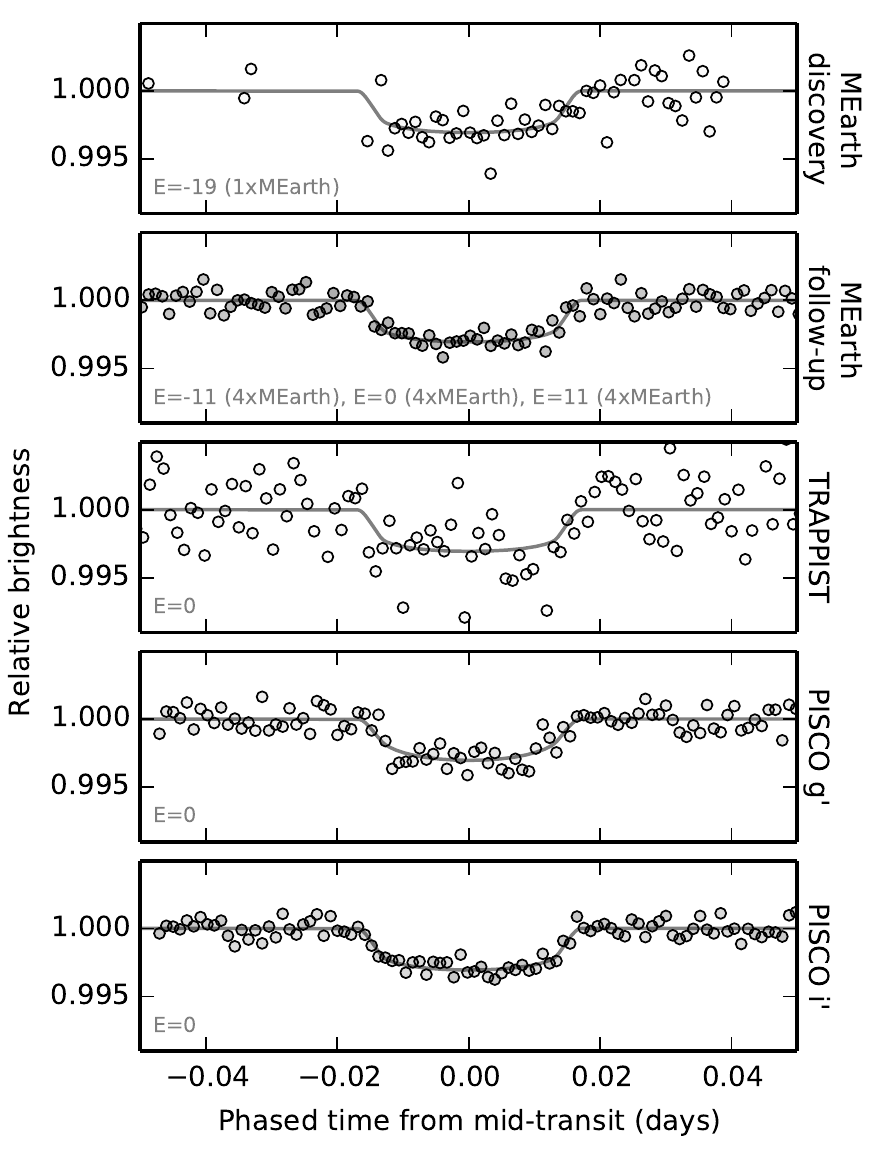} 
   \caption{{\bf Photometric measurements of transits of GJ 1132b.} Light curves from the MEarth-South, TRAPPIST and PISCO telescopes/imagers were fitted with a transit model (grey lines) and a Gaussian process noise model (subtracted from this plot), and averaged to 1.5-min bins for visual clarity. For MEarth-South, both the initial triggered `discovery' observations and the subsequent `follow-up' observations are shown. Labels indicate the transit event (with $E$ as an integer number of planetary periods) and, for MEarth-South, the number of telescopes used. The opacities of binned points are inversely proportional to their assigned variances, representing their approximate weights in the model fit. The raw data and details of the fit are presented in Methods; $g'$ and $i'$ refer to the wavelength bandpasses used from the PISCO imager.}
   \label{f:tlc}
\end{figure}  
\pagebreak
A search of extant data (4,208 observations over 333 nights) for periodic signals revealed a 1.6-day candidate that included this event and reached a detection statistic\ci{13} of 9.1$\sigma$. Follow-up photometry of subsequent predicted transits with four MEarth-South telescopes, the TRAPPIST (TRAnsiting Planets and PlanetesImals Small Telescope) telescope\ci{14}, and the PISCO (Parallel Imager for Southern Cosmology Observations) multiband imager\ci{15} on the Magellan Clay telescope confirmed the transit signal as being consistent with a planet-sized object blocking 0.26\% of the star's light. We began precise Doppler monitoring with the HARPS (High Accuracy Radial Velocity Planet Searcher) spectrograph\ci{16} on 6 June 2015 and gathered 25 radial velocity measurements for determining the planetary mass (Fig. 2).

\begin{figure}
   \centering
   \includegraphics[width=4.125in]{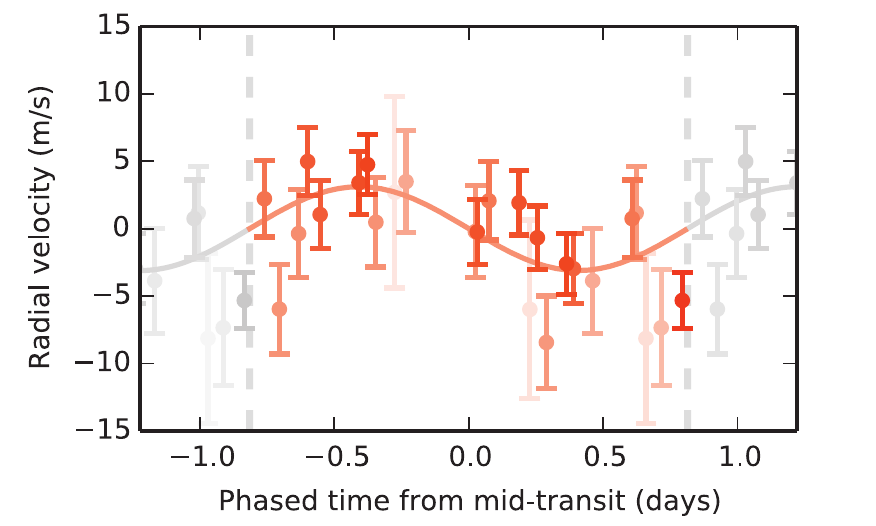} 
   \caption{{\bf Radial velocity changes over the orbit of GJ 1132b.} Measurements of the star's line-of-sight velocity, taken by the HARPS spectrograph, are shown phased to the planetary orbital period determined from the light curves (orange points, with duplicates shown in grey). Error bars correspond to $1\sigma$. The darkness of each point is proportional to its weight in the model fit, which is the inverse of its variance as predicted by a radial velocity noise model. For a circular orbit, the star's reflex motion to the planet has a semi-amplitude of $K_{\star} = 2.76 \pm 0.92 {\rm~m~s}^{-1}$.}
   \label{f:rvc}
\end{figure}

The distance to GJ 1132 has been measured through trigonometric parallax to be $12.04\pm0.24$ parsecs\ci{17}, a value that we independently validate with MEarth astrometry (see Methods). Together with empirical relations among the intrinsic luminosities, masses and radii of M-dwarf stars\ci{18,19}, the parallax enables us to estimate the mass and radius of GJ 1132. These estimates are not biased by physically associated luminous companions, which are ruled out by published photometry results and the HARPS spectra. Likewise, unassociated background stars are too faint in archival imaging at the current sky position of this high-proper-motion star to corrupt our estimates of the stellar parameters. Table 1 presents the physical properties of the star (GJ 1132) and planet (GJ 1132b), combining the inferred stellar properties with analyses of the transit light curves (Fig. 1) and radial velocity observations (Fig. 2). The radius of the planet is 40\% that of GJ 1214b (ref. 20), a well studied mini-Neptune exoplanet that orbits with a similar period around a similar host star.

\begin{table}
\caption{}
\begin{center}
\begin{tabular}{lc}
{\bf Table 1: System properties for GJ 1132b} & \\
\hline
Parameter	 & Value \\
\hline
~\\
{\bf Stellar parameters} \\
Photometry 								&	$V = 13.49, J = 9.245, K = 8.322$ 	\\
Distance to star, $D_{\star}$					&	$12.04 \pm 0.24$ parsecs 		\\
Mass of star, $M_{\star}$						&	$0.181 \pm 0.019 M_{\odot}$		\\
Density of star, $\rho_{\star}$					&	$29.6 \pm 6.0 {\rm~g~cm}^{-3}$ 			\\
Radius of star, $R_{\star}$						&	$0.207 \pm 0.016 R_{\odot}$		\\
Luminosity of star, $L_{\star}$					&	$0.00438 \pm 0.00034 L_{\odot}$	\\
Effective temperature, $T_{\rm eff}$				&	$3270 \pm 140$ K 				\\
Metallicity, [Fe/H]							&	$-0.12 \pm 0.15$				\\
Age of star, $\tau_{\star}$						&	$>5$ Gyr						\\
~\\
{\bf Transit and radial velocity parameters} \\
Orbital period, $P$ (days)						&	$1.628930 \pm 0.000031$		\\
Time of mid-transit, $t_{0}$ (BJD$_{\rm TDB}$; days)	&	$2457184.55786 \pm 0.00032$		\\
Eccentricity, $e$							&	$0$ (fixed)					\\
Planet-to-star radius ratio, $R_{\rm p}/R_{\star}$		&	$0.0512 \pm 0.0025$			\\
Scaled orbital distance, $a/R_{\star}$			&	$16.0 \pm 1.1$					\\
Impact parameter, $b$						&	$0.38 \pm 0.14$				\\
Radial velocity semi-amplitude, $K_{\star}$		&	$2.76 \pm 0.92 {\rm~m~s}^{-1}$		\\
Systemic velocity, $\gamma_{\star}$				&	$+35 \pm 1 {\rm~km~s}^{-1}$		\\
~\\
{\bf Planet parameters} \\
Radius of planet, $R_{\rm p}$						&	$1.16 \pm 0.11 R_{\oplus}$		\\	
Mass of planet, $M_{\rm p}$						&	$1.62 \pm 0.55 M_{\oplus}$		\\	
Density of planet, $\rho_{\rm p}$					&	$6.0 \pm 2.5 {\rm~g~cm}^{-3}$		\\
Surface gravity on planet, $g_{\rm p}$				&	$1170 \pm 430 {\rm~cm~s}^{-2}$		\\
Escape velocity, $V_{\rm esc}$					&	$13.0 \pm 2.3 {\rm~km~s}^{-1}$		\\
Equilibrium temperature, $T_{\rm eq}$			&								\\	
~~~~assuming Bond albedo of 0.00				&	$579 \pm 15$ K				\\
~~~~assuming Bond albedo of 0.75				&	$409 \pm 11$ K					\\
~\\
\hline

\end{tabular}
\end{center}
\label{default}
{\small  
Transit and radial velocity parameters were estimated from a Markov chain Monte Carlo (MCMC) analysis, including an external constraint on the stellar density when deriving $P$, $t_{0}$, $R_{\rm p}/R_\star$, $a/R_\star$, and $b$ (see Methods). Planetary properties were derived from the combined stellar, transit, and radial velocity parameters. $L_\odot$, luminosity of the Sun; $M_\odot$, mass of the Sun; $R_\odot$, radius of the Sun; ${\rm BJD_{TDB}}$, Barycentric Julian Date in the Barycentric Dynamical Time system; $a$, orbital semimajor axis; $M_\oplus$, mass of Earth; $R_\oplus$, radius of Earth. 
}

\end{table}%

~

GJ 1132b's average density resembles that of the Earth, and is well matched by a rock/iron bulk composition. A theoretical mass-radius curve\ci{21} for a two-layer planet composed of 75\% magnesium silicate and 25\% iron (by mass) is consistent with our estimates for GJ 1132b (Fig. 3). This model assumes that the core is pure iron, the mantle is pure magnesium silicate, and the interior contains no water\ci{21}. These simplifications mean that the iron fraction should not be taken as absolute; the model simply represents a characteristic mass-radius locus that matches Earth and Venus. This same composition also matches the masses and radii of Kepler-78b (refs 8, 9), Kepler-10b (ref. 7), Kepler-93b (ref. 10), Kepler-36b (ref. 6), CoRoT-7b (ref. 5), and HD 219134b (ref. 11) to within $1\sigma$. All of these planets are smaller than 1.6 Earth radii, a transition radius above which most planets require thick hydrogen/helium envelopes to explain their densities\ci{22}. At the $1\sigma$ lower bound of GJ 1132b's estimated mass, models\ci{23} indicate that replacing only 0.2\% of the rock/iron mix with a hydrogen/helium layer would increase the planet's radius to 1.4 times that of the Earth, substantially larger than the observed value. Detection of GJ 1132b's mass is currently only at the $3\sigma$ level, but continued Doppler monitoring will shrink the 35\% mass uncertainty and enable more detailed comparison with other planets and compositional models. 

We searched for additional planets both as other transits in the MEarth-South light curve and as periodic signals in the HARPS residuals. Although we made no notable discoveries, we highlight that compact, coplanar, multiple-planet systems are common around small stars\ci{24,25}. Further exploration of the GJ 1132 system could reveal more, potentially transiting, planets.

As a relatively cool rocky exoplanet with an equilibrium temperature between 580 K (assuming a Bond albedo of 0) and 410 K (assuming a Venus-like Bond albedo of 0.75), GJ 1132b may have retained a substantial atmosphere. At these temperatures, the average thermal speeds of atoms or molecules heavier than helium are less than one-eighth of the escape velocity, suggesting an atmosphere could be stable against thermal escape. This is not the case for the other rocky exoplanets for which precise densities are known, all of which are considerably hotter. The rocky planet Kepler-78b (refs 8, 9), which is comparable in size and density to GJ 1132b, receives 200 times more irradiation than GJ 1132b. Whether the atmosphere of GJ 1132b was initially dominated by hydrogen/helium-rich gas accreted from the primordial nebula or by volatiles outgassed from the planetary interior, its composition probably evolved substantially over the age of the system, which we estimate to exceed 5 billion years (gigayears, Gyr) (see Methods). Irradiated well beyond the runaway greenhouse limit\ci{26}, surface water would extend up to high altitudes where it could be destroyed by photolysis and its hydrogen rapidly lost to space. When the star was young and bright at ultraviolet wavelengths, an atmosphere with high concentrations of water could lose hydrogen at the diffusion limit, of the order of 10$^{13}$ atoms per cm$^{2}$ per second or 10 Earth oceans per gigayear. Depending on surface weathering processes, the oxygen left behind might persist as O$_{2}$ in the atmosphere\ci{26,27}. In this scenario, water would constitute a trace component in an atmosphere otherwise dominated by O$_{2}$, N$_{2}$, and CO$_{2}$. However, large uncertainties in the size of the initial hydrogen reservoir, in the history of the star's ultraviolet luminosity, in the contribution of late volatile delivery, and in the evolutionary effect of the system's likely spin-orbit synchronization preclude firm {\em a priori} statements about the composition of the atmosphere.

\begin{figure*}[t] 
   \centering
   \includegraphics[width=6.5in]{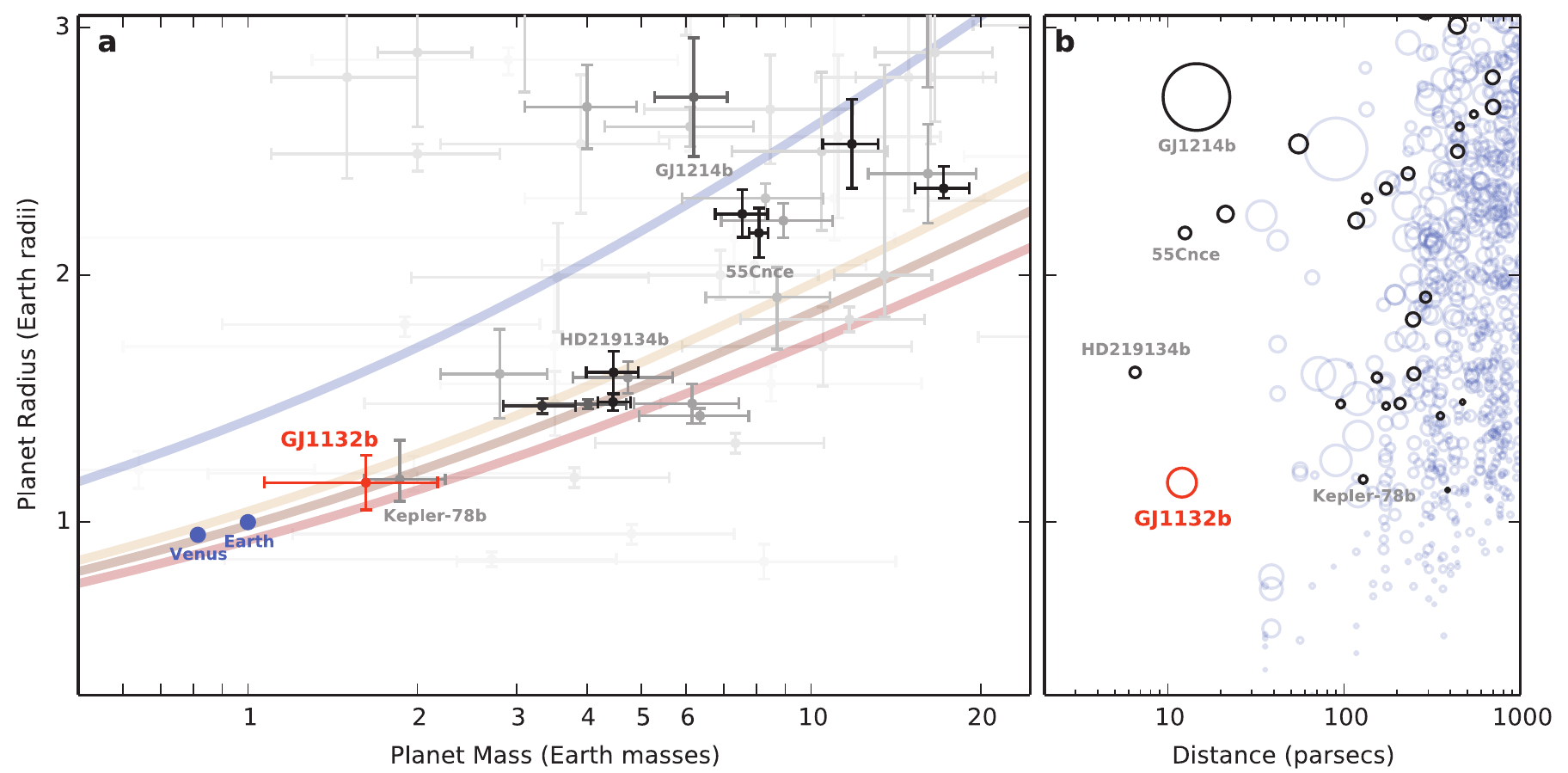} 
   \caption{{\bf Masses, radii, and distances of known transiting planets.} {\bf a}, The radius and mass of GJ 1132b (orange) are shown, along with those of other exoplanets (grey). Also shown are mass-radius curves predicted by theoretical models\ci{21} for planets composed of 100\% H$_2$O (blue line), and for two-component planets composed of MgSiO$_{3}$ on top of Fe cores that are 0\% (light brown), 25\% (darker brown) or 50\% (red) of the total mass. Planets with smaller fractional mass and radius uncertainties are darker. {\bf b}, Symbol area is proportional to transit depth. In comparison with other transiting exoplanets, those with masses detected at $>2.5\sigma$ (black) and those without masses detected at such level (blue), GJ 1132b is the most accessible terrestrial planet for spectroscopic observations of its atmosphere, owing to the proximity and small size of its parent star.}
   \label{f:context}
\end{figure*}

Future spectroscopic investigation of the planetary atmosphere will be enabled by the proximity and small radius of the star. When viewed in transmission during transit, one scale height of an O$_{2}$-rich atmosphere would overlap 10 parts per million (p.p.m.) of the stellar disk. For comparison, a 60-orbit Hubble Space Telescope transmission spectrum of GJ 1214b achieved a transit depth precision of 25 p.p.m. in narrow wavelength bins\ci{28}. Deeper Hubble observations of GJ 1132, which is 50\% brighter than GJ 1214, would have the potential to detect molecular absorption features in GJ 1132b's atmosphere. Observations with the James Webb Space Telescope (JWST), set to launch in 2018, could measure the transmission spectrum over a broader wavelength range and require less telescope time. The long-wavelength capabilities of JWST may also allow it to detect the thermal emission from the planet; such emission represents 40-130 p.p.m. of the system flux at a wavelength of 10 $\mu$m, and 160-300 p.p.m. of the flux at 25 $\mu$m (for the range of albedos considered above). Provided that the planet is not too cloudy, combined transmission and emission spectra could ascertain the abundances of strongly absorbing molecular species. If such constraints on the dominant infrared opacity sources can be obtained, observations of the planet's thermal phase curve would be sensitive to complementary information, including the total atmospheric mass\ci{29,30}. Such observations will inform our understanding of how the strong tides and intense stellar activity of the M-dwarf planetary environment influence the evolution of terrestrial atmospheres. This understanding will be important for the long-term goal of looking for life on planets orbiting nearby small stars.

{\small \it Received 3 August; accepted 23 September 2015; doi:10.1038/nature15762.}

\subsection{References}
{\footnotesize
	\begin{itemize}
\setlength\itemsep{.2em}

\item[]	1.	Dressing, C. D. \& Charbonneau, D. The occurrence rate of small planets around small stars. {\em Astrophys. J.} 767, 95 (2013).  \\
\item[]	2.	Morton, T. D. \& Swift, J. The radius distribution of planets around cool stars. {\em Astrophys. J.} 791, 10 (2014).  \\
\item[]	3.	Dressing, C. D. \& Charbonneau, D. The occurrence of potentially habitable planets orbiting M dwarfs estimated from the full Kepler dataset and an empirical measurement of the detection sensitivity. {\em Astrophys. J.} 807, 45 (2015).  \\
\item[]	4.	Muirhead, P. S. et al. Characterizing the cool KOIs. III. KOI 961: a small star with large proper motion and three small planets. {\em Astrophys. J.} 747, 144 (2012).  \\
\item[]	5.	Haywood, R. D. et al. Planets and stellar activity: hide and seek in the CoRoT-7 system. {\em Mon. Not. R. Astron. Soc.} 443, 2517-2531 (2014).  \\
\item[]	6.	Carter, J. A. et al. Kepler-36: a pair of planets with neighboring orbits and dissimilar densities. {\em Science} 337, 556-559 (2012). \\
\item[]	7.	Dumusque, X. et al. The Kepler-10 planetary system revisited by HARPS-N: a hot rocky world and a solid Neptune-mass planet. {\em Astrophys. J.} 789, 154 (2014).  \\
\item[]	8.	Pepe, F. et al. An Earth-sized planet with an Earth-like density. {\em Nature} 503, 377-380 (2013).  \\
\item[]	9.	Howard, A. W. et al. A rocky composition for an Earth-sized exoplanet. {\em Nature} 503, 381-384 (2013).  \\
\item[]	10.	Dressing, C. D. et al. The mass of Kepler-93b and the composition of terrestrial planets. {\em Astrophys. J.} 800, 135 (2015).  \\
\item[]	11.	Motalebi, F. et al. The HARPS-N rocky planet search I. HD 219134 b: a transiting rocky planet in a 4 planet system at 6.5 pc from the Sun. {\em Astron. Astrophys.} (in the press). Preprint at \url{http://adslabs.org/adsabs/abs/2015arXiv150708532M/}.  \\ 
\item[]	12.	Irwin, J. M. et al. The MEarth-North and MEarth-South transit surveys: searching for habitable super-Earth exoplanets around nearby M-dwarfs. {\em Proc. 18th Conf. Cambridge Work on Cool Stars, Stellar Systems, \& Sun} (Eds van Belle, G. \& Harris, H. C.) 767-772 \url{http://adslabs.org/adsabs/abs/2015csss...18..767I/} (2015). 
\item[]	13.	Berta, Z. K., Irwin, J., Charbonneau, D., Burke, C. J. \& Falco, E. E. Transit detection in the MEarth survey of nearby M dwarfs: bridging the clean-first, search-later divide. {\em Astron. J.} 144, 145 (2012).  \\
\item[]	14.	Gillon, M. et al. TRAPPIST: a robotic telescope dedicated to the study of planetary systems. {\em EPJ Web Conf.} 11, 06002 (2011). \\
\item[]	15.	Stalder, B. et al. PISCO: the Parallel Imager for Southern Cosmology Observations. In {\em Proc. SPIE} (eds Ramsay, S. K., McLean, I. S. \& Takami, H.) Vol. 9147, 91473Y (2014).
\item[]	16.	Mayor, M. et al. Setting new standards with HARPS. {\em Messenger} 114, 20-24 (2003).  \\
\item[]	17.	Jao, W.-C. et al. The solar neighborhood XIII: parallax results from the CTIOPI 0.9-m program: stars with $\mu >= 1$"/year (MOTION sample). {\em Astron. J.} 129, 1954 (2005).  \\
\item[]	18.	Delfosse, X. et al. Accurate masses of very low mass stars: IV. Improved mass-luminosity relations. {\em Astron. Astrophys.} 364, 217-224 (2000). \\
\item[]	19.	Hartman, J. D. et al. HATS-6b: a warm Saturn transiting an early M dwarf star, and a set of empirical relations for characterizing K and M dwarf planet hosts. {\em Astron. J.} 149, 166 (2015).  \\
\item[]	20.	Charbonneau, D. et al. A super-Earth transiting a nearby low-mass star. {\em Nature} 462, 891-894 (2009).  \\
\item[]	21.	Zeng, L. \& Sasselov, D. D. A detailed model grid for solid planets from 0.1 through 100 Earth masses. {\em Publ. Astron. Soc. Pacif.} 125, 227-239 (2013).  \\
\item[]	22.	Rogers, L. A. Most 1.6 Earth-radius planets are not rocky. {\em Astrophys. J.} 801, 41 (2015).  \\
\item[]	23.	Lopez, E. D. \& Fortney, J. J. Understanding the mass-radius relation for sub-Neptunes: radius as a proxy for composition. {\em Astrophys. J.} 792, 1 (2014).  \\
\item[]	24.	Ballard, S. \& Johnson, J. A. The Kepler dichotomy among the M dwarfs: half of systems contain five or more coplanar planets. Preprint at \url{http://adslabs.org/adsabs/abs/2014arXiv1410.4192B/} (2014).
\item[]	25.	Muirhead, P. S. et al. Kepler-445, Kepler-446 and the occurrence of compact multiples orbiting mid-M dwarf stars. {\em Astrophys. J.} 801, 18 (2015).  \\
\item[]	26.	Kasting, J. F., Whitmire, D. P. \& Reynolds, R. T. Habitable zones around main sequence stars. {\em Icarus} 101, 108-128 (1993).  \\
\item[]	27.	Luger, R. \& Barnes, R. Extreme water loss and abiotic O2 buildup on planets throughout the habitable zones of M dwarfs. {\em Astrobiology} 15, 119-143 (2015).  \\
\item[]	28.	Kreidberg, L. et al. Clouds in the atmosphere of the super-Earth exoplanet GJ 1214b. {\em Nature} 505, 69-72 (2014).  \\
\item[]	29.	Selsis, F., Wordsworth, R. \& Forget, F. Thermal phase curves of nontransiting terrestrial exoplanets 1. Characterizing atmospheres. {\em Astron. Astrophys.} 532, A1 (2011).  \\
\item[]	30.	Koll, D. D. B. \& Abbot, D. S. Deciphering thermal phase curves of dry, tidally locked terrestrial planets. {\em Astrophys. J.} 802, 21 (2015).  \\
\end{itemize}}

{\footnotesize

\noindent {\bf Supplementary Information} is available in the online version of the paper.

~

\noindent {\bf Acknowledgements} We thank the staff at the Cerro Tololo Inter-American Observatory for assistance in the construction and operation of MEarth-South; J. Winn and J. Berta-Thompson for comments on the manuscript; S. Seager and A. Zsom for conversations that improved the work; L. Delrez for her independent analysis of the TRAPPIST data; and J. Eastman, D. Dragomir and R. Siverd for their efforts to observe additional transits. The MEarth Project acknowledges funding from the David and Lucile Packard Fellowship for Science and Engineering, and the National Science Foundation, and a grant from the John Templeton Foundation. The opinions expressed here are those of the authors and do not necessarily reflect the views of the John Templeton Foundation. The development of the PISCO imager was supported by the National Science Foundation. HARPS observations were made with European Space Observatory (ESO) Telescopes at the La Silla Paranal Observatory. TRAPPIST is a project funded by the Belgian Fund for Scientific Research, with the participation of the Swiss National Science Foundation. Z.K.B.-T. is funded by the MIT Torres Fellowship for Exoplanet Research. X.B., X.D., T.F. and A.W. acknowledge the support of the French Agence Nationale de la Recherche and the European Research Council.  M.G. and E.J. are FNRS Research Associates. V.N. acknowledges a CNPq/BJT Post-Doctorate fellowship and partial financial support from the INCT INEspaço. N.C.S. acknowledges the support from the Portuguese National Science Foundation (FCT) as well as the COMPETE program.

~

\noindent  {\bf Author Contributions} The MEarth team (D.C., J.I., Z.K.B.-T., E.R.N. and J.A.D.) discovered the planet, organized the follow-up observations, and led the analysis and interpretation. Z.K.B.-T. analysed the light curve and radial velocity data, and wrote the manuscript. J.I. designed, installed, maintains, and operates the MEarth-South telescope array, identified the first triggered transit event, and substantially contributed to the analysis and interpretation. D.C. leads the MEarth Project, and assisted in analysis and writing the manuscript. E.R.N. determined the metallicity, kinematics, and rotation period of the star. J.A.D. confirmed the star's trigonometric parallax and helped install the MEarth-South telescopes. The HARPS team (N.A.-D., X.B., F.B., X.D., T.F., C.L., M.M., V.N., F.P., N.C.S., S.U. and A.W.) obtained spectra for Doppler velocimetry, with N.A.-D. and X.B. leading the analysis of those data. M.G. and E.J. gathered photometric observations with TRAPPIST. A.A.S. and B.S. gathered photometric observations with PISCO. All authors read and discussed the manuscript.

~

\noindent {\bf Author Information} Reprints and permissions information is available at \url{http://www.nature.com/reprints}. The authors declare no competing financial interests. Readers are welcome to comment on the online version of the paper. Correspondence and requests for materials should be addressed to Z.K.B.-T. (\url{zkbt@mit.edu}).

}

\renewcommand{\figurename}{{\bf Extended Data Figure}}
\renewcommand{\thefigure}{{\bf \arabic{figure}}}
\setcounter{figure}{0}

\section{Methods}
\paragraph{Distance to the star}
GJ 1132's coordinates are 10:14:51.77 -47:09:24.1 (International Celestial Reference System, epoch 2000.0), with proper motions of (-1046; 416) milliarcseconds (mas) per year and a trigonometric parallax of $\pi = 83.07 \pm 1.69$ mas, as determined by the RECONS (Research Consortium on Nearby Stars) survey\ci{17}. We qualitatively confirm this parallax with independent observations from MEarth-South, using analyses like those that have been applied to the northern survey\ci{31}. The motion of GJ 1132 relative to background stars in MEarth-South imaging closely matches the prediction made by the RECONS parallax (Extended Data Fig. 1). We do not quote the value of $\pi$ derived from MEarth-South because we have not yet cross-validated the astrometric performance of the system against other measurements. Literature photometric observations of GJ 1132 include photoelectric photometry ($U = 16.51 \pm 0.03$, $B = 15.17 \pm 0.03$)\ci{32}, charge-coupled device (CCD) photometry ($V = 13.49 \pm 0.03$, $RC = 12.26 \pm 0.02$, $I_{C} = 10.69 \pm 0.02$)\ci{17}, 2MASS near-infrared photometry ($J = 9.245 \pm 0.026$, $H = 8.666 \pm 0.027, K_{s} = 8.322 \pm 0.027$)\ci{33}, and WISE infrared photometry ($W1 = 8.170 \pm 0.023$, $W2 = 8.000 \pm 0.020$, $W3 = 7.862 \pm 0.018$, $W4 = 7.916 \pm 0.184$). The colour ($V-K_{s} = 5.168 \pm 0.040$) and absolute magnitude ($M_{V} = 13.088 \pm 0.054$) of GJ 1132 are consistent with those of single M4V dwarfs\ci{34}. 
\begin{marginfigure}[-10cm]
      \hspace{-0.5cm}
   \includegraphics[width=2in]{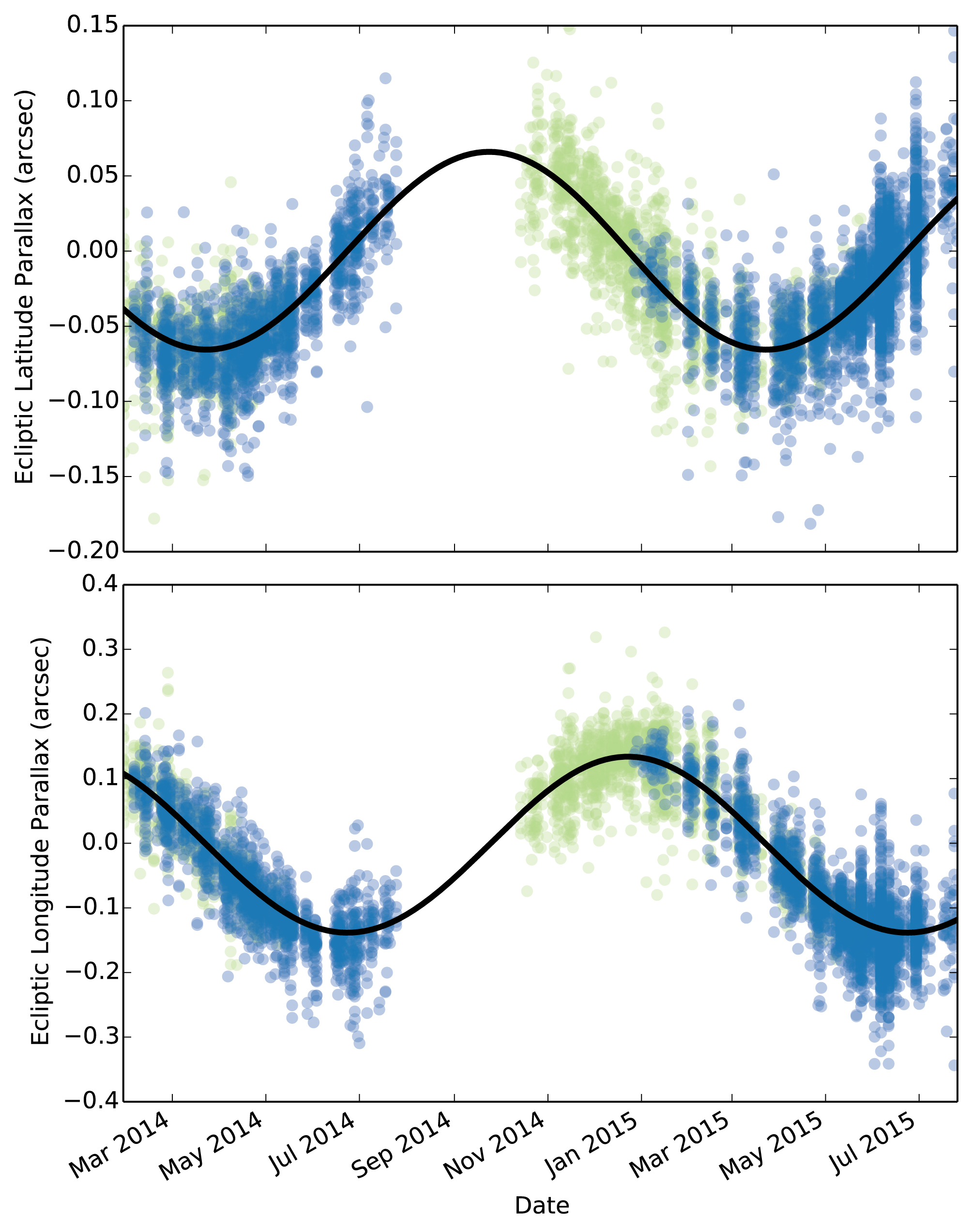}
   \caption{{\bf Astrometry of GJ 1132 from MEarth-South.}  Measurements of the star GJ 1132's position in MEarth-South images, along the directions of ecliptic latitude (top) and longitude (bottom). As described elsewhere\ci{31}, a fitted offset between data gathered at a field rotation of $0^\circ$ (blue) and $180^\circ$ (green) has been removed. The published RECONS proper motion\ci{17} has been subtracted, and a model fixed to the published 83.07 mas parallax (black line) closely matches the MEarth-South observations.}
   \label{f:tlc}
\end{marginfigure}  
\paragraph{Metallicity of the star}
Before discovering the planet, we gathered a near-infrared spectrum of GJ 1132 with the FIRE spectrograph on the Magellan Baade telescope. We shifted the spectrum to a zero-velocity wavelength scale\ci{35}, measured equivalent widths and compared the spectra by eye with solar metallicity spectral type standards\ci{35}. The spectrum indicates a near-infrared spectral type of M4V-M5V (Extended Data Fig. 2), slightly later than the optical spectral of M3.5V listed in the PMSU (Palomar/Michigan State University) catalogue\ci{36}. Using the measured equivalent width of the K-band sodium feature ($4.7$\AA) and an empirical calibration\ci{35} that has been corrected for its known temperature dependence\ci{37}, we estimate the stellar metallicity to be [Fe/H] $= -0.12 \pm 0.15$ and quote this value in Table 1. For comparison, a relation using additional spectral regions and calibrated for stars of GJ 1132's spectral type and earlier\ci{38} also yields [Fe/H] = -0.1, while one for GJ 1132's spectral type and later\ci{39} yields [Fe/H] = -0.2 (both with uncertainties of about 0.15 dex).

\begin{figure}[h]
   \centering
   \includegraphics[width=4.125in]{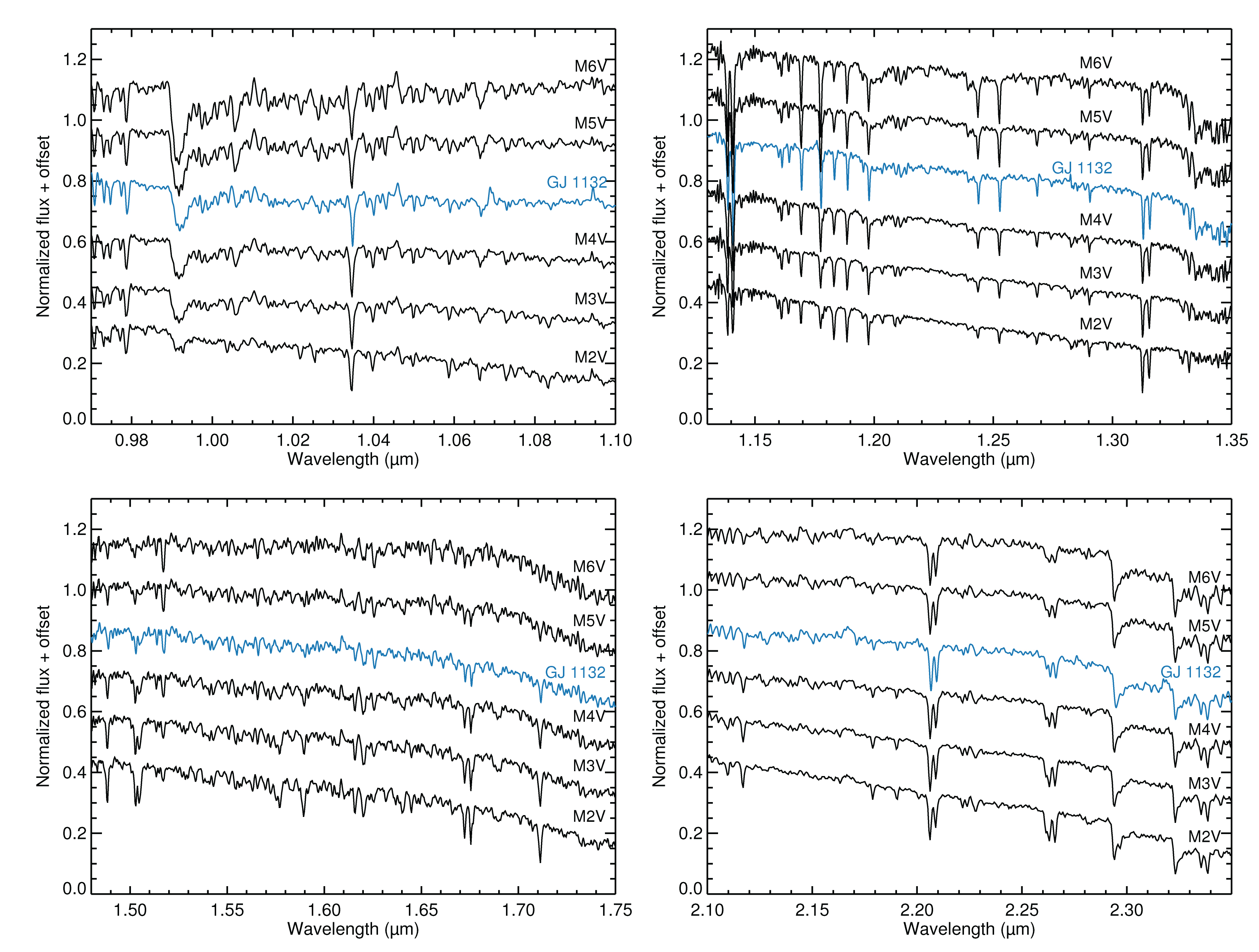}
   \caption{{\bf Near-infared spectrum of GJ 1132.} Observations of GJ 1132's spectrum obtained with the FIRE spectrograph on the Magellan Baade telescope are compared in the $z$, $J$, $H$ and $K$ telluric windows (left to right, top to bottom) to the solar-metallicity composite spectral type standards from ref. 35. The FIRE spectra have been smoothed to match the $R = 2,000$ resolution of the standards. GJ 1132's near-infrared spectral type is M4V-M5V.}
   \label{f:tlc}
\end{figure}  
\vspace{-.5cm}
\paragraph{Mass of the star}
Dynamical mass measurements of M-dwarfs show that tight relationships exist between near-infrared absolute magnitudes and stellar mass\ci{18}. We use these calibrations to calculate masses from the $J$, $H$ and $K$ magnitudes (after converting between the 2MASS and CIT photometric systems). Taking the mean of these masses and adopting an uncertainty that is the quadrature sum of the 2.7\% error propagated from the measurement uncertainties and the 10\% scatter we assume for the relations, we adopt a stellar mass of $M_{\star} = 0.181 \pm 0.019 M_{\odot}$, where $M_{\star}$ is the mass of the star and $M_{\odot}$ is the mass of the Sun.

\paragraph{Radius of the star}
From this mass, we use an empirical $M_\star-\rho_\star$ relation\ci{19} calibrated to eclipsing binary systems to estimate a density of $\rho_{\star} = 29.6 \pm 6.0 {\rm~g~cm}^{-3}$, corresponding to $R_\star = 0.207 \pm 0.016 R_\odot$ for GJ 1132. We adopt those values, noting that they agree with two other mass-radius relations: the radius predicted by long-baseline optical interferometry of single stars\ci{40} is $R_\star = 0.211 \pm 0.014 R_\odot$, and that by the Dartmouth evolutionary models\ci{41} is $R_\star = 0.200 \pm 0.016 R_\odot$ (for [Fe/H] = -0.1, assuming a uniform prior on age between 1 Gyr and 10 Gyr). The quoted errors do not include an assumed intrinsic scatter in any of the mass--radius relations, but the consistency among the three estimates suggests that any contribution from scatter would be smaller than the uncertainty propagated from the stellar mass. 

\paragraph{Bolometric luminosity of the star}
We combine the parallax and photometry with bolometric corrections to determine the total luminosity of GJ 1132, testing three different relations to estimate bolometric corrections from colour. The Mann et al. relation\ci{42} between $BC_{V}$ and $V-J$ colour yields a bolometric luminosity of $0.00402 L_\odot$. The Leggett et al. relation\ci{43} between $BC_{K}$ and $I-K$ colour yields $0.00442 L_\odot$. The Pecaut and Mamajek compilation of literature bolometric corrections\ci{44}, when interpolated in $V-K_{s}$ colour to determine $BC_{V}$, yields $0.00469 L_\odot$. We adopt the mean of these three values, with an uncertainty that is the quadrature sum of the systematic error (the 6.3\% standard deviation of the different estimates) and the uncertainty propagated from the measurement uncertainties (about 5\% in all three cases), as our final estimate of the bolometric luminosity: $L_\star = 0.00438 \pm 0.00034 L_\odot$. From this, we calculate the stellar effective temperature as $T_{\rm eff} = 5772 {\rm K} \times (L_\star/L_\odot)^{1/4} \times (R_\star/R_\odot)^{-1/2} = 3270 \pm 140 {\rm K}$. The luminosity and temperature we infer\ci{37} from the FIRE spectra ($L_\star = 0.0044 \pm 0.001 L_\odot$, $T_{\rm eff} = 3130 \pm 120 {\rm K}$) are consistent with the quoted values.

\begin{marginfigure}[5cm]
   \hspace{-0.5cm}
   \includegraphics[width=2in]{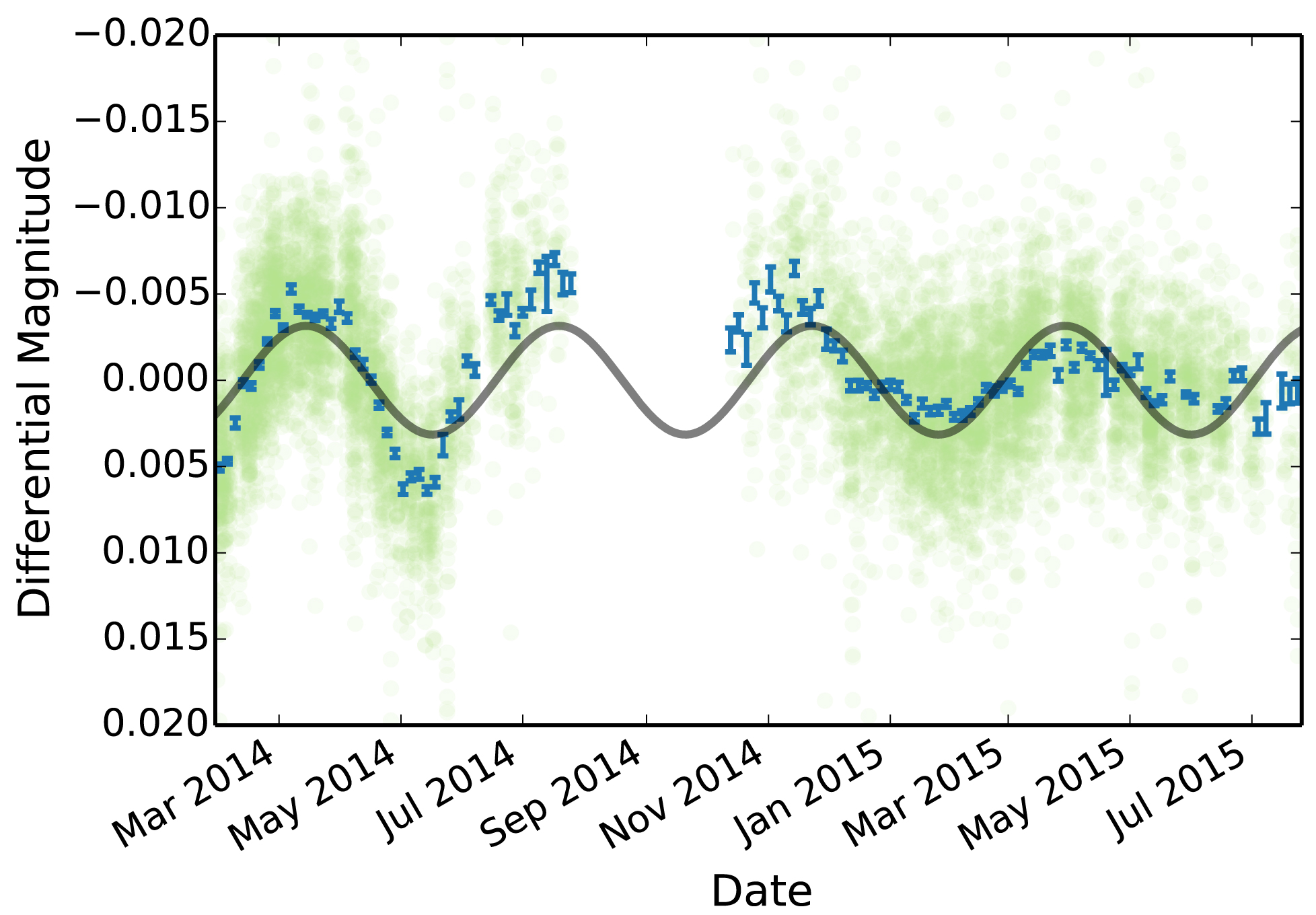}
   \caption{{\bf Photometric starspot modulations of GJ 1132.} MEarth-South photometry (with dots representing single pointings and error bars representing $\pm1\sigma$  uncertainty ranges on weighted averages over four-day bins) probes starspots that are rotating in and out of view, and indicates that GJ 1132 has a rotation period of approximately 125 days. The rotational modulation was identified using a methodology similar to that used in previous MEarth work\ci{48}.}
   \label{f:tlc}
\end{marginfigure}  

\marginnote{\tiny \color{gray} $^*$$UVW$ velocities have been updated to use the proper motion stated above. The values given in the published article were based on the proper motion from ref. 64, and were also with respect to the local standard of rest. We adopt the solar velocities of ref. 65.}
\paragraph{Age of the star}
GJ 1132's motion through the Galaxy of ($U_{\rm LSR}$, $V_{\rm LSR}$, $W_{\rm LSR}$) = (-47, -32, -2) km s$^{-1}$ is consistent with a kinematically older stellar population$^*$. M4 dwarfs tend to show strong H$\alpha$ emission for about 4 Gyr\ci{45}; the lack of H$\alpha$ emission in the HARPS spectrum indicates that GJ 1132 is probably older than that. The star instead shows weak H$\alpha$ absorption, which is an indicator of non-zero magnetic activity in stars as cool as this\ci{46}. We detect emission in the Ca II H line, with a weak intensity that is comparable to that of Barnard's Star and other slowly rotating stars in the HARPS M-dwarf sample\ci{47}. Applying published methods\ci{48} to the MEarth-South photometry, we measure a rotation period of 125 days for GJ 1132 (Extended Data Fig. 3). M-dwarfs spin more slowly as they age, with less massive stars reaching longer rotation periods at old ages\ci{48}. Therefore, the rotation period suggests that the system could be almost as old as Barnard's Star (0.16 $M_\odot$), which has a 130-day rotation period\ci{49} and is 7-13 Gyr old\ci{50}. GJ 1132's age is probably comparable to or greater than that of Proxima Centauri (0.12 $M_\odot$), which has an 83-day period\ci{49,51} and is 5-7 Gyr old, assuming it to be coeval with the gyrochronologically and asteroseismically age-dated $\alpha$ Centauri system\ci{52}. From these comparisons, we conclude that GJ 1132 is probably older than 5 Gyr.

\paragraph{Assumption of a circular orbit}
Calculating the timescale for tidal circularization\ci{53} as $t_{\rm circ} = \frac{2PQ}{63\pi} \times \frac{M_{\rm p}}{M_\star} \times \left( \frac{a}{R_{\rm p}} \right)^{5}$, assuming a tidal quality factor of $Q = 100$ appropriate for rocky exoplanets, yields $4 \times 10^{5}$ years for GJ 1132b. This is much shorter than the age of the system; therefore, we assume the eccentricity to be negligible and fix it to 0 in our estimation of the other planetary properties. Perturbations from other (undetected) planets in the system could potentially induce a small equilibrium eccentricity.

\paragraph{Photometric observations and analysis}
The high-cadence light curves analysed here were gathered over four nights using six telescopes on three mountains. The triggered event (event $E = -19$, relative to the ephemeris in Table 1) was collected on MEarth-South telescope 3. At the time of this triggered detection, 296 stars in the MEarth-South sample had been observed at least 100 times, but only nine of these stars had been observed as thoroughly as GJ 1132. After identification of the periodic signal, we obtained follow-up photometry with MEarth-South telescopes 2, 3, 4 and 8 on the nights of 23 May 2015, 10 June 2015 and 28 June 2015 ($E = -11$, $0$ and $11$). Exposure times were 18 s, yielding a 47-s cadence for the high-cadence light curves. We observed the transit on the night of 10 June 2015 ($E = 0$) with the TRAPPIST telescope, in a wide $I+z$ bandpass, with exposure times of 10 s and a cadence of 21 s. We also observed this transit with the PISCO multiband imager, which was installed on the Magellan Baade telescope and undergoing a commissioning run, taking 10-s exposures simultaneously in $g'$, $r'$, $i'$, $z'$ bandpasses at a cadence of 36 s. We actively defocused Baade, resulting in donut-shaped point spread functions spreading each star's light over about 200 pixels. The PISCO $r'$ and $z'$ light curves exhibited unexplained systematics and were not used for further analysis.

To the high-cadence light curves, we fitted a transit model\ci{54} with the following parameters: orbital period $P$, time of mid-transit $t_{0}$, planet-to-star radius ratio $R_{\rm p}/R_\star$, star-to-orbit radius ratio $R_\star/a$, impact parameter $b$, and separate baseline flux levels for each transit observed on each telescope. The PISCO and TRAPPIST photometry showed trends that correlated with airmass, so we also included coefficients for a trend linear with airmass as free parameters. Limb-darkening was treated with a quadratic approximation, using fixed coefficients\ci{55} calculated from a PHOENIX atmosphere model for a 3,300 K, [Fe/H] = 0, $\log g = 5$ star. These coefficients ($u1, u2$) were: (0.1956, 0.3700) for MEarth and TRAPPIST (both approximated as $\frac{1}{3}i' + \frac{2}{3}z'$), (0.4790, 0.3491) for PISCO $g'$ (approximated as Sloan $g'$), and (0.2613,0.3737) for PISCO $i'$ (approximated as Sloan $i'$). We performed a Levenberg-Marquardt maximization of the posterior probability of this model, rejected $>4\sigma$ outliers among the data, repeated the maximization, and increased the per-point uncertainty estimates until each transit exhibited a reduced $\chi^{2} \le 1$. This resulted in typical per-point uncertainties of 2.6 (MEarth), 3.6 (TRAPPIST), 1.9 (PISCO $g'$), and 1.2 (PISCO $i'$) mmag. Most light curves showed some evidence for time-correlated noise in their residuals. To marginalize over the uncertainty introduced by these correlations, we use the Gaussian process regression package George\ci{56} to model each set of transit residuals as a Gaussian process with non-zero covariance between the datapoints\ci{57}. We use a Matérn-3/2 kernel function to describe this covariance as a function of separation in time, and include two free parameters per transit: $\log t_{\rm gp}$ (where $t_{\rm gp}$ is the correlation timescale) and $\log a_{\rm gp}$ (where $a_{\rm gp}$ is an amplitude). We mapped the probability distribution of the model parameters using the emcee implementation\ci{58} of an affine-invariant Markov chain Monte Carlo (MCMC) sampler\ci{59}, assuming flat priors on all parameters. Extended Data Fig. 4 shows raw light curves of individual transits, along with model curves sampled from the posterior.

The ingress time ($\tau = 3.7 \pm 1.0$ min from first to second contact) is measured imprecisely in this fit, compared to the total transit duration ($T_{\rm tot} = 47.0 \pm 1.4$ min from first to fourth contact). Systematic astrophysical errors (incorrect limb-darkening, starspots, unidentified transit timing variations) can bias light-curve estimates of the ingress time, and therefore also the parameters $R_\star/a$ and $b$ that depend strongly on $\tau$. Therefore, we estimate the planet properties without relying on this ingress measurement, by including external constraints on both the stellar mass and the stellar density. We sample values of $M_\star$ and $\rho_\star$ from Gaussian distributions ($0.181 \pm 0.019 M_\odot$ and $29.6 \pm 6.0 {\rm~g~cm}^{-3}$), use Kepler's Third Law to compute $R_\star/a$ from the stellar density and the period, and then calculate $b$ from the transit duration and these new $R_\star/a$ samples. We quote the marginalized values and uncertainties for these and other light-curve parameters in Table 1. This procedure forces consistency between the light-curve fit and the inferred stellar properties and shifts the estimates from $R_\star/a = 0.0738 \pm 0.0092$ and $b = 0.58 \pm 0.14$ (inferred from the light curves alone) to the values in Table 1. The effect on $P$, $t_{0}$, or $R_{\rm p}/R_\star$ is negligible. If we ignore the external $\rho_\star$ information, the inferred stellar and planetary radii would be consistent with those in Table 1 but with larger error bars ($R_\star = 0.243 \pm 0.031 R_\odot$ and $R_{\rm p} = 1.37 \pm 0.22 R_\oplus$).

\begin{figure*}
   \includegraphics[width=5.5in]{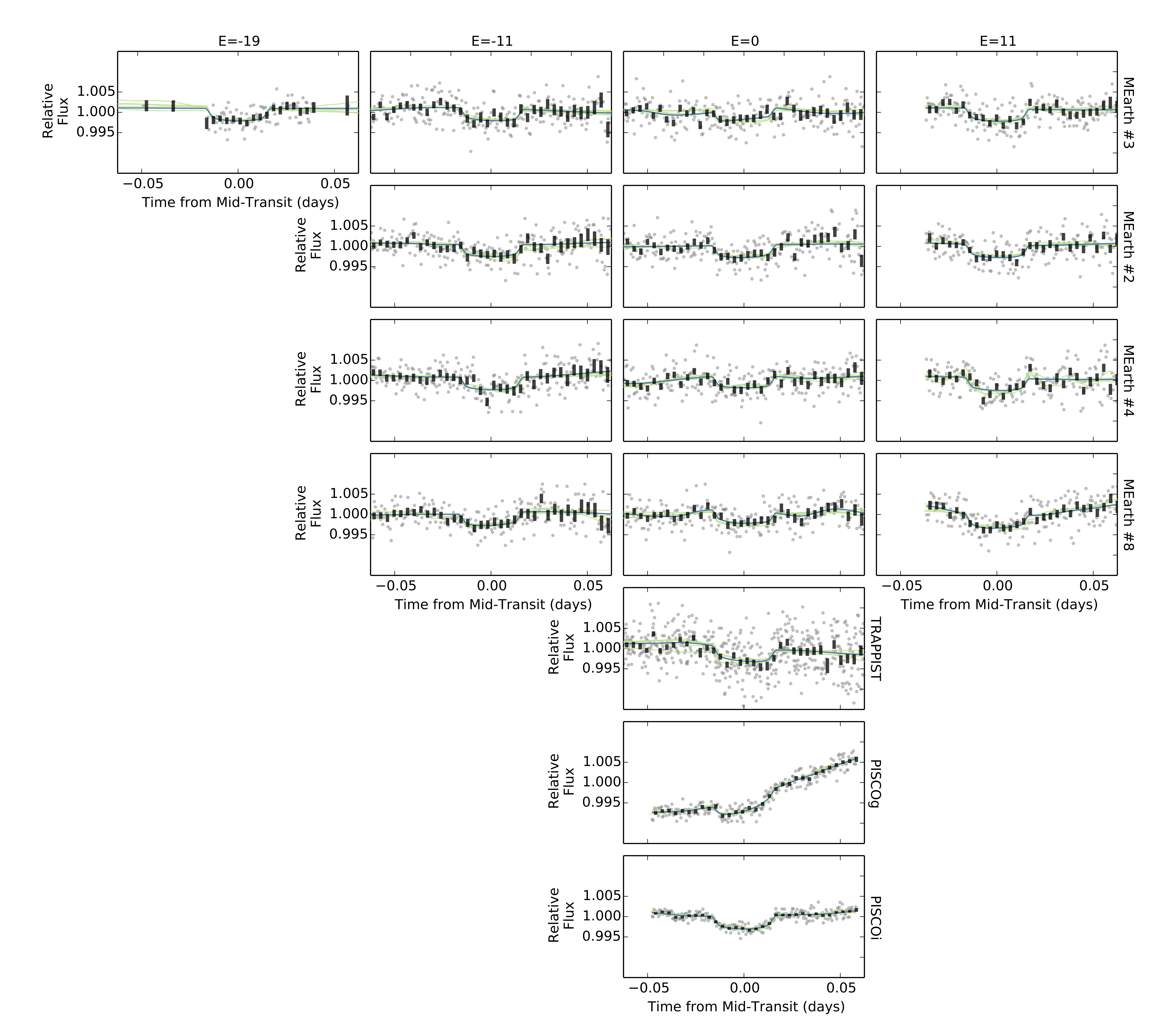}
   \vspace{-4cm} \caption{{\bf Raw transit light curves of GJ 1132b.} Light curves are shown both unbinned (grey points) and in five-minute bins (black bars, representing the $\pm1\sigma$  uncertainty range for the weighted average in each bin), and separated by telescope (row) and transit event (column). Model curves are shown, with the Gaussian process noise model conditioned on the observations, for parameters sampled from the posterior (green) and for the maximum likelihood parameters (blue). This is the complete set of light-curve data behind the transit parameter fits.}

   \label{f:tlc}
\end{figure*}  
    \vspace{4cm}

\paragraph{Radial velocity observations and analysis}
We first gathered reconnaissance spectra of GJ 1132 with the CHIRON\ci{60} spectrograph on the SMARTS 1.5-metre telescope at CTIO. Once these spectra ruled out large radial-velocity variations corresponding to binary star companions, we began observations with the HARPS spectrograph on the La Silla 3.6-metre telescope. We gathered 25 HARPS spectra of GJ 1132 between 6 June 2015 and 18 July 2015, using an exposure time of 40 min. The spectra span 380 nm to 680 nm in wavelength. For each observation, we constructed a comparison template by co-adding all the other spectra and measured the relative radial velocity as the shift required to minimize the $\chi^2$ of the difference between the spectrum and this template\ci{61,62}. Telluric lines were masked using a template of lines made with a different, much larger, data set. We see no evidence for rotational broadening of the spectra, indicating that the projected rotational velocity, $v\sin i$, is less than $2 {\rm~km~s}^{-1}$. The median of the internal radial velocity uncertainty estimates\ci{61} is $3 \rm{~m~s}^{-1}$ per observation.

Over the timespan of our observations, these radial velocities are dominated by the orbital motion of the known transiting planet, rather than by other planets, blended stellar binaries, or stellar activity. The tallest peak in the periodogram of the radial velocities is consistent with the transit-derived period for GJ 1132b. The line bisectors show no significant correlation with the measured velocities (Pearson correlation $p$-value of 0.27) and no clear periodicities. The H$\alpha$ equivalent width and the full-width half-maximum of the HARPS cross-correlation function are not correlated with the velocities (Pearson correlation $p$-values of 0.72 and 0.57).

We fit the velocities with a model corresponding to a circular orbit for the planet, with flat priors on three free-floating parameters: the radial velocity semi-amplitude $K_\star$, the systemic velocity $\gamma_\star$, and a stellar jitter term $\sigma_{\star, \rm jitter}$ that is added in quadrature to the internal velocity uncertainties. We include all terms that depend on $\sigma_{\star, \rm jitter}$ in the likelihood, as well as the usual $\chi^2$ term. $P$ and $t_{0}$ were fixed to the values determined from the transit analysis. We use emcee\ci{58} to sample from the posterior probability distributions of these parameters, and quote marginalized uncertainties on $K_\star$ in Table 1. In this fit, the value of $K_\star$ is $>0$ in 99.7\% of the MCMC samples. The inferred value of $\sigma_{\star, \rm jitter}$ is smaller than the individual uncertainties ($<1.9 {\rm~m~s^{-1}}$ at 68\% confidence), and a good fit is obtained if $\sigma_{\star, \rm jitter}$ is fixed to 0 ($\chi^2 = 26.67$ for 23 degrees of freedom). The $\chi^2$ of a fit where both $K_\star$ and $\sigma_{\star, \rm jitter}$ are fixed to 0 is significantly worse (38.56 for 24 degrees of freedom). As a check, we repeated the MCMC fit allowing the phase ($t_{0}$) to float. This radial velocity fit predicted the known mid-transit times to within $1.5\sigma$ ($3.0 \pm 1.9$ h after the true transit times). Relaxing the assumption of a circular orbit yields inferred distributions of ($e\cos \omega$, $e\sin \omega$) that are consistent with (0, 0) but substantially increases the uncertainty on the semi-amplitude ($K_\star = 3.6 \pm 2.4 {\rm~m~s}^{-1}$). With future Doppler measurements, it will be possible to measure the planet's eccentricity and independently test the assumption we make here that tides damped the eccentricity to small values.

\paragraph{Transiting exoplanet population comparison}
Figure 3 compares GJ 1132b to other known transiting exoplanets. Data for this plot were drawn from the NASA Exoplanet Archive\ci{63} on 25 August 2015. Egregious errors in the catalogue were replaced with appropriate literature values. In the mass-radius panel, individual planets are shaded with a white-to-black greyscale that is inversely proportional $\left(\sigma_{M}/M_{\rm p}\right)^{2} + \left(\sigma_{R}/R_{\rm p}\right)^{2}$, where $\sigma_{M}$ and $\sigma_{R}$ are the uncertainties on the planet mass, $M_{\rm p}$, and radius, $R_{\rm p}$. Planets with precise measurements are dark, and those with large mass or radius uncertainties are light. In the distance-radius panel, some planets did not have distances listed in the Exoplanet Archive. For those systems, we calculated approximate distances from the $J$-band magnitudes, estimated stellar radii, effective temperatures, and a table of bolometric corrections\ci{44} (interpolating in $T_{\rm eff}$ to estimate $BC_{J}$).

\paragraph{Code availability}
Analyses were conducted primarily in Python. Although not cleanly packaged for general use, for the sake of transparency we make the custom code used for transit and radial velocity fitting available at \url{http://github.com/zkbt/transit}. It relies on three freely available packages: eb (\url{http://github.com/mdwarfgeek/eb}), emcee (\url{http://github.com/dfm/emcee}), and George (\url{http://github.com/dfm/george}). Code used to generate the exoplanet population comparison is available at \url{http://github.com/zkbt/exopop}.

\subsection{Additional References}

{\footnotesize
	\begin{itemize}
\setlength\itemsep{.2em}

\item[] 31.	Dittmann, J. A., Irwin, J. M., Charbonneau, D. \& Berta-Thompson, Z. K. Trigonometric parallaxes for 1507 nearby mid-to-late M dwarfs. {\em Astrophys. J.} 784, 156 (2014).  \\
\item[] 32.	Eggen, O. J. Catalogs of proper-motion stars. I. Stars brighter than visual magnitude 15 and with annual proper motion of 1 arcsec or more. {\em Astrophys. J. Suppl.} 39, 89 (1979).  \\
\item[] 33.	Skrutskie, M. F. et al. The two micron all sky survey (2MASS). {\em Astron. J.} 131, 1163-1183 (2006).  \\
\item[] 34.	Winters, J. G. et al. The solar neighborhood. XXXV. Distances to 1404 M dwarf systems within 25 pc in the southern sky. {\em Astron. J.} 149, 5 (2014).  \\
\item[] 35.	Newton, E. R. et al. Near-infrared metallicities, radial velocities, and spectral types for 447 nearby M dwarfs. {\em Astron. J.} 147, 20 (2014).  \\
\item[] 36.	Hawley, S. L., Gizis, J. E. \& Reid, N. I. The Palomar/MSU nearby star spectroscopic survey. II. The southern M dwarfs and investigation of magnetic activity. {\em Astron. J.} 113, 1458 (1997).  \\
\item[] 37.	Newton, E. R., Charbonneau, D., Irwin, J. \& Mann, A. W. An empirical calibration to estimate cool dwarf fundamental parameters from H-band spectra. {\em Astrophys. J.} 800, 85 (2015). \\
\item[] 38.	Mann, A. W., Brewer, J. M., Gaidos, E., Lépine, S. \& Hilton, E. J. Prospecting in late-type dwarfs: a calibration of infrared and visible spectroscopic metallicities of late K and M dwarfs spanning 1.5 dex. {\em Astron. J.} 145, 52 (2013).  \\
\item[] 39.	Mann, A. W. et al. Prospecting in ultracool dwarfs: measuring the metallicities of mid- and late-M dwarfs. {\em Astron. J.} 147, 160 (2014).  \\
\item[] 40.	Boyajian, T. S. et al. Stellar diameters and temperatures. II. Main-sequence K- and M-stars. {\em Astrophys. J.} 757, 112 (2012).  \\
\item[] 41.	Dotter, A. et al. The Dartmouth stellar evolution database. {\em Astrophys. J. Suppl.} 178, 89-101 (2008).  \\
\item[] 42.	Mann, A. W., Feiden, G. A., Gaidos, E., Boyajian, T. \& von Braun, K. How to constrain your M dwarf: measuring effective temperature, bolometric luminosity, mass, and radius. {\em Astrophys. J.} 804, 64 (2015).  \\
\item[] 43.	Leggett, S. K., Allard, F., Geballe, T. R., Hauschildt, P. H. \& Schweitzer, A. Infrared spectra and spectral energy distributions of late M and L dwarfs. {\em Astrophys. J.} 548, 908-918 (2001).  \\
\item[] 44.	Pecaut, M. J. \& Mamajek, E. E. Intrinsic colors, temperatures, and bolometric corrections of pre-main-sequence stars. {\em Astrophys. J. Suppl.} 208, 9 (2013).  \\
\item[] 45.	West, A. A. et al. Constraining the age-activity relation for cool stars: the Sloan digital sky survey data release 5 low-mass star spectroscopic sample. {\em Astron. J.} 135, 785-795 (2008).  \\
\item[] 46.	Walkowicz, L. M. \& Hawley, S. L. Tracers of chromospheric structure. I. Observations of Ca II K and H$\alpha$ in M dwarfs. {\em Astron. J.} 137, 3297-3313 (2009).  \\
\item[] 47.	Bonfils, X. et al. The HARPS search for southern extra-solar planets. XXXI. The M-dwarf sample {\em Astron. Astrophys}. 549, A109  (2013).  \\ 
\item[] 48.	Irwin, J. et al. On the angular momentum evolution of fully-convective stars: rotation periods for field M-dwarfs from the MEarth transit survey. {\em Astrophys. J.} 727, 56 (2010). \\
\item[] 49.	Benedict, G. F. et al. Photometry of Proxima Centauri and Barnard's Star using Hubble space telescope fine guidance sensor 3: a search for periodic variations. {\em Astron. J.} 116, 429-439 (1998).  \\
\item[] 50.	Feltzing, S. \& Bensby, T. The galactic stellar disc. {\em Phys. Scr.} 2008, T133 (2008).  \\
\item[] 51.	Kiraga, M. \& Stepien, K. Age-rotation-activity relations for M dwarf stars. {\em Acta Astron.} 57, 149-172 (2007). \\
\item[] 52.	Mamajek, E. E. \& Hillenbrand, L. A. Improved age estimation for solar-type dwarfs using activity-rotation diagnostics. {\em Astrophys. J.} 687, 1264 (2008). \\
\item[] 53.	Goldreich, P. \& Soter, S. Q in the solar system. {\em Icarus} 5, 375-389 (1966).  \\
\item[] 54.	Mandel, K. \& Agol, E. Analytic light curves for planetary transit searches. {\em Astrophys. J.} 580, L171-L175 (2002). \\
\item[] 55.	Claret, A., Hauschildt, P. H. \& Witte, S. New limb-darkening coefficients for PHOENIX/1D model atmospheres. {\em Astron. Astrophys.} 546, A14 (2012).  \\
\item[] 56.	Ambikasaran, S., Foreman-Mackey, D., Greengard, L. \& Hogg, D. W. \& O'Neil, M. Fast direct methods for Gaussian processes and the analysis of NASA Kepler mission data. Preprint at \url{http://arxiv.org/abs/1403.6015} (2014).
\item[] 57.	Gibson, N. P. et al. A Gaussian process framework for modelling instrumental systematics: application to transmission spectroscopy. {\em Mon. Not. R. Astron. Soc.} 419, 2683-2694 (2012).  \\
\item[] 58.	Foreman-Mackey, D., Hogg, D. W., Lang, D. \& Goodman, J. Emcee: the MCMC hammer. {\em Publ. Astron. Soc. Pacif.} 125, 306-312 (2013).  \\
\item[] 59.	Goodman, J. \& Weare, J. Ensemble samplers with affine invariance. {\em Comm. App. Math. Comp. Sci.} 5, 65-80 (2010).  \\
\item[] 60.	Tokovinin, A. et al. CHIRON: a fiber fed spectrometer for precise radial velocities. {\em Publ. Astron. Soc. Pacif.} 125, 1336-1347 (2013).  \\
\item[] 61.	Bouchy, F., Pepe, F. \& Queloz, D. Fundamental photon noise limit to radial velocity measurements. {\em Astron. Astrophys.} 374, 733-739 (2001).  \\
\item[] 62.	Astudillo-Defru, N. et al. The HARPS search for southern extra-solar planets. {\em Astron. Astrophys.} 575, A119 (2015).  \\
\item[] 63.	Akeson, R. L. et al. The NASA exoplanet archive: data and tools for exoplanet research. {\em Publ. Astron. Soc. Pacif.} 125, 989-999 (2013).  \\
\item[] 64. Bakos, G. A. et al. Revised Coordinates and Proper Motions of the Stars in the Luyten Half-Second Catalog {\em Astrophys. J. Suppl.} 141, 1, 187-193 (2002). \\
\item 65. Sch\"onrich, R. et al. Local kinematics and the local standard of rest. {\em Mon. Not. R. Astron. Soc.} 403, 4, 1829-1833 (2010). \\
	\end{itemize}
}

\end{document}